\documentstyle[aps,prl,graphicx,floats]{revtex}
\newcommand{\beq}{\begin{equation}}
\newcommand{\eeq}{\end{equation}}

\begin{document}

\baselineskip=18pt

\begin{center}
{\Large\bf  Mixed initial conditions to estimate the dynamic critical
exponent in short-time Monte Carlo simulation}

\vskip 1.1cm
{\bf Roberto da Silva\footnote{E-mail: rsilva@dfm.ffclrp.usp.br},~~
     Nelson A. Alves\footnote{E-mail: alves@quark.ffclrp.usp.br}~and~
   J.R. Drugowich de Fel\'{\i}cio\footnote{E-mail: drugo@usp.br}}
\vskip 0.1cm
{\it Departamento de F\'{\i}sica e Matem\'atica, FFCLRP
     Universidade de S\~ao Paulo. Av. Bandeirantes 3900. \\
     CEP 014040-901 \, Ribeir\~ao Preto, SP, Brazil}

\vskip 0.3cm
\today
\vskip 0.4cm
\end{center}
\begin{abstract}
 We explore the initial conditions in short-time critical dynamics to propose
a new method to evaluate the dynamic exponent $z$. Estimates are obtained
with high precision for 2D Ising model and 2D Potts
model for three and four states by performing heat-bath Monte Carlo
simulations.


\end{abstract}
\vskip 0.1cm
{\it PACS-No.: 64.60.Fr, 64.60.Ht, 02.70.Lq, 75.10.Hk}



\nopagebreak

\indent

The description of static critical phenomena in terms of finite size scaling
(FSS) relations, developed by Fisher {\it et al.} \cite
{Fisher72,FisherLectures} has been extended by Halperin, Hohenberg, Ma and
Suzuki \cite{HH1,Suzuki} to include dynamical {\it properties} of the
system. Later, Janssen {\it et al.} \cite{Janssen1}, and independently Huse 
\cite{Huse} found evidence for an universal behavior far from equilibrium.

As discussed by Janssen {\it et al.} 
one finds an universal behavior already in the early stages of the
relaxation process for systems prepared at an initial state characterized by
non-equilibrium values of the order parameter. As a consequence, they could
advance the existence of a new critical exponent $\theta $, independent of
the known set of static exponents and of the dynamic critical exponent $z$.
This new exponent characterizes the so called ``critical initial slip'', the
anomalous increasing of the magnetization when the system is quenched to the
critical temperature $T_{c}$.

That new universal stage has been exhaustively investigated to confirm
theoretical predictions and to enlarge our knowledge on phase transitions
and critical phenomena. In this sense, several models and algorithms \cite
{Universal97,Sch97,Zhang99} have been used, as toy models, in order to check
the ability of the new approach in obtaining dynamic and static critical
exponents. Results are in good agreement with pertinent results for static
exponents and seems to be confident even for the new critical exponent $%
\theta $. However, a reliable technique to obtain the dynamic exponent $z$
is lacking. A first proposal by Li {\it et al.} \cite{Li95} using a
time-dependent Binder's cumulant yields estimates with low precision when
compared with other techniques \cite{Grassberger,Nightingale}. An
alternative way which uses another kind of cumulant, proposed by Zheng \cite
{Review}, gives the right answer for the 2D Ising model but fails in
determining the value of $z$ for the 3-state Potts model \cite{Review} and
for the Ising model with multispin interactions \cite{Simoes}.

In this letter, we introduce and check a new technique to obtain the
exponent $z$, combining the behavior of the order parameter and its second
moment when the system is submitted to different initial conditions.

Before presenting our proposal, we shall review the main results in short
time dynamics. 

Although Halperin {\it et al.} \cite{HH1} have studied
systems with different dynamics, we consider only systems without
conservation laws, the so called Model A \cite{HH2} because our discussion
aims dynamics generated by heat bath dynamics. Therefore, we consider a
magnetic system prepared at high temperature ($T>>T_{c}$) with a small
nonzero magnetization $m_{0}$, (this can be achieved with a small external
magnetic field $h$) and quenched to the critical temperature $T_{c}$ without
any external magnetic field. If the system is allowed to relax towards
equilibrium with the dynamics of model A, the magnetization obeys the
following scaling relation (generalized to $k{\rm th}$ moment), 
\begin{equation}
M^{(k)}(t,\tau ,L,m_{0})=b^{-k\beta /\nu }M^{(k)}(b^{-z}t,b^{1/\nu }\tau
,b^{-1}L,b^{x_{0}}m_{0})\,.  \label{magk}
\end{equation}
Here $b$ is an arbitrary spatial scaling factor, $t$ is the time evolution
and $\tau $ is the reduced temperature, $\tau =(T-T_{c})/T_{c}$. The
exponents $\beta $ and $\nu $ are the static critical exponents, while $z$
is the dynamic one. $M^{(k)}=\left\langle M^{k}\right\rangle $ are the $k%
{\rm th}$ moments of magnetization. This scaling relation depends on the
initial magnetization $m_{0}$ and gives origin to a new, independent
critical exponent $x_{0}$, the scaling dimension of the initial
magnetization, which is related to $\theta $. 

From Eq. (\ref{magk}) we can derive the power law increasing of the
magnetization, observed in the initial stage of the dynamic relaxation. For
this purpose, we consider large lattice sizes $L$ at $\tau =0$ with $%
b=t^{1/z}$. This leads to the scaling relation 
\begin{equation}
M(t,m_{0})=t^{-\beta /\nu z}M(1,t^{x_{0}/z}m_{0})  \label{mag0}
\end{equation}
for the first moment of the magnetization $M^{(1)}\equiv M$. By expanding
this equation for small $m_{0}$, we have the following power law, 
\begin{equation}
M(t)\sim m_{0}t^{\theta }\,,  \label{m0}
\end{equation}
where, as anticipated, we identify $\theta =(x_{0}-\beta /\nu )/z$. Here we
also have the condition that $t^{x_{0}/z}m_{0}$ is small, which sets a time
scale $t_{0}\sim m_{0}^{-z/x_{0}}$ \cite{Janssen1,Foundations} where that
phenomena can be observed.

On the other hand, it has been realized the existence of another important
dynamic process from an initial ordered state \cite
{Stauffer92,Stauffer93,Sch96}, which represents another fixed point in the
context of renormalization group approach. This leads to a different universal
behavior of the dynamic relaxation process also described by Eq.~(\ref{magk}%
) with $m_{0}=1$. In particular, dealing with large enough lattice sizes at
the critical temperature ($\tau =0$), one obtains a power law decay of the
magnetization 
\begin{equation}
M(t)\sim t^{-\beta /\nu z}\,,  \label{m1}
\end{equation}
which follows from Eq.~(\ref{magk}) when we choose $b^{-z}t=1$ and shows the
average magnetization is not zero as would be expected from a disordered
initial state.

Equation (\ref{magk}) and their particular forms in (\ref{m0}) and (\ref{m1}%
) can be used to determine relations involving static critical exponents and
the dynamic exponent $z$ \cite{Review,Foundations}.

The observables in short-time analysis are described by different scaling
relations according to the initial magnetization. In particular, the second
moment $M^{(2)}(t,L)$, 
\begin{equation}
M^{(2)}=\left\langle \left( \frac{1}{L^{d}}\sum\limits_{i=1}^{N}\sigma
_{i}\right) ^{2}\right\rangle =\frac{1}{L^{2d}}\left\langle
\sum\limits_{i=1}^{N}\sigma _{i}^{2}\right\rangle +\frac{1}{L^{2d}}%
\sum\limits_{i\neq j}^{N}\left\langle \sigma _{i}\sigma _{j}\right\rangle \,,
\label{second}
\end{equation}
behaves as $L^{-d}$ since in the short-time evolution with initial condition 
$m_{0}=0$ the spatial correlation length is very small compared with the
lattice size $L$. Thus we have \cite{Sch97,Review} 
\begin{equation}
M^{(2)}(t,L)=t^{-2\beta /\nu z}\,M^{(2)}(1,t^{-1/z}L)\sim t^{(d-2\beta /\nu
)/z}\,.  \label{increasemag2}
\end{equation}
Moreover, under this initial condition one can also define the
time-dependent Binder's cumulant at the critical temperature, 
\begin{equation}
U(t,L)=1-\frac{M^{(4)}(t,L)}{3(M^{(2)}(t,L))^{2}}\,,  \label{binder0}
\end{equation}
which leads at $T=T_{c}$ to the FSS relation 
\begin{equation}
U(t,L)=U(b^{-z}t,b^{-1}L)\,,  \label{r13}
\end{equation}
and the exponent $z$ can be independently evaluated through scaling
collapses for different lattice sizes \cite{Li95,Li96}.

In order to obtain more precise estimates for the dynamic exponent $z$,
another cumulant has been proposed \cite{Review}. It is given by 
\begin{equation}
U_{2}(t,L)=\frac{M^{(2)}(t,L)}{(M(t,L))^{2}}-1  \label{binder1}
\end{equation}
and should behave as 
\begin{equation}
U_{2}(t)\sim t^{d/z}  \label{binder11}
\end{equation}
when one starts from an ordered state. In this case, curves for all the
lattices lay on the same straight line without any re-scaling in time and
results in more precise estimates for $z$. However, application of this
procedure has not been successful in at least two well known models: the
two-dimensional $q=3$ Potts model \cite{Review} and the Ising model with
three spin interactions in just one direction \cite{Simoes}. The reason for
the above disagreement could be related to the value of the second
term of r.h.s in Eq.~(\ref{second}) when $m_{0}=1$.

A plausible way to circumvent this problem is to work with different initial
conditions. For this we decided to follow the evolution of the ratio $%
F_{2}=M^{(2)}/M^{2}$ but using different initial conditions to calculate
each one of the mean values. The reason is we know the behavior of the
second moment of the magnetization when samples are initially disordered $%
(m_{0}=0)$ and also the dependence on time of the magnetization of samples
initially ordered $(m_{0}=1)$. Under the above mentioned conditions the
ratio behaves as 
\begin{equation}
F_{2}(t,L)=\frac{\left. M^{(2)}(t,L)\right| _{m_{0}=0}}{\left.
(M(t,L))^{2}\right| _{m_{0}=1}}\sim \frac{t^{(d-2\beta /\nu )/z}}{t^{-2\beta
/\nu z}}=t^{d/z}\,,  \label{binder2}
\end{equation}
which has the same potential law that the cumulant mentioned before but
requires two independent simulations instead of one used for calculating $%
U_{2}$.

In short-time Monte Carlo (MC) simulations, the time scale $t$ is settled in
units of whole lattice updates, and does not depend on the initial
conditions. However,  the same dynamics (Metropolis, Glauber or heat-bath)
should be used in both simulations \cite{Universal97,Sch97}.

Now we present our estimates obtained for 2D Ising model, $q=3$ and $q=4$
Potts models.

We have performed independent heat-bath (HB) MC simulations for a large
lattice $(L=144)$ according to the required initial conditions and the
evolutions have been done until the maximum time $t=200$ MC sweeps with $%
N=10000$ samples. This gives averages for the magnetization and its second
moment. We have performed this kind of simulation 20 times to obtain our
final estimates for each moment in function of $t$. Since all runs are
independent due to the choice of random numbers, we have a total of 
400 time series for Eq.~(\ref{binder2}).

Our results are shown on a log-log scale for the time interval [10, 100] in
Fig.1 and Fig.2, respectively for three and four-state Potts model (a
similar figure is obtained for 2D Ising model) and they appear to be clearly
consistent with a power law in that time interval. The error bars are also
presented in those figures but they are hardly seen in that scale.

Our estimates for $z$ comes from a least square fitting in the time interval 
$[t_{i},t_{f}]$. Due to our small statistical errors, we can make a
systematic study for the range in $t$ where we find acceptable
goodness-of-fit $Q$ \cite{Press}. As examples, we obtain for 2D Ising model, 
$z=2.1435(2)$ in the time interval $[10,200]$ with $Q=10^{-250}$, $%
z=2.1359(3)$ in $[50,200]$, with $Q=10^{-41}$, and the most acceptable $%
Q=0.99$ in $[30,90]$, which yields $z=2.1565(7)$. This value is presented in
Table 1, where we also include some estimates for comparison. Here we
observe that our result is in agreement with more recent estimates within two
standard-deviations.
 This indicates our statistical errors are presumably underestimated possibly
due to corrections to scaling. 

We complete the overview in Table 1 with data from \cite{Wang97}. Estimates
have been obtained from the long time behavior of the magnetization for the
square lattice (sq), $z=2.168(5)$, for the triangle (TP), $z=2.180(9)$, and
for the honeycomb (hc) lattice, $z=2.167(8)$, while from damage spreading in
short-time the authors quote $2.166(7)$, $2.164(7)$ and $2.170(10)$ for sq,
TP and hc lattices.

For $q=3$ Potts model, our study monitoring $Q$ gives $z=2.198(2)$ in the
interval [50, 90] with $Q=0.82$. This result agrees with $z=2.203(11)$ (Ref. 
\cite{Review}) obtained from the Binder cumulant in Eq.~(\ref{binder0}), and
with the estimate obtained in \cite{Sch97} with HB algorithm from the second
moment $M^{(2)}(t)$ in short-time analysis.

The value in \cite{Zhang99} refers to TP lattice, presenting further
numerical evidence (comparing \cite{Sch97} and \cite{Zhang99}) to the
dynamic universality. Reference \cite{Review} also presents the value $%
2.14(3)$, obtained from Eq.~(\ref{binder1}), in clear disagreement with $%
2.203(11)$ as commented in \cite{Review}. On the other hand, our estimate
from Eq.~(\ref{binder2}) gives a value in full agreement with the Binder
cumulant and collapse data analysis in short time.

The case $q=4$ has been less studied. Our analysis gives $z=2.290(3)$ in the
interval $[60, 90]$ with $Q=0.72$. Here, we stress the importance of
monitoring $Q$, since we may find values for $z$ as large as $z=2.3483(2)$
in [10, 200] but with unacceptable value, $Q=10^{-269}$ or $z=2.3532(3)(2)$
in [10, 100] with $Q= 10^{-202}$.

As a final comment, the suspected $z$ as been weakly \cite{Arcangelis} or
even independent on $q$ \cite{Bonfim} is not supported by the most recently
results presented in Table 1.


\newpage
\vspace{0.4cm} 

{\bf Acknowledgements} 
\vspace{0.4cm}

R. da Silva gratefully acknowledges support by FAPESP (Brazil), and 
N. Alves by CNPq (Brazil). Thanks are also due to DFMA for computer 
facilities at IFUSP.


\vspace{1.0cm}

{\huge Figure Captions:} \\
\begin{description}
\item[Figure 1.] Time evolution of $F_2(t)$ for 2D three-state Potts model.\\
\item[Figure 2.] Time evolution of $F_2(t)$ for 2D four-state Potts model. 
\end{description}


\begin{table}[ht]
\caption{\baselineskip=0.8cm Dynamic exponent $z$ for 2D Ising model, three
and four-state Potts model.}\renewcommand{\tablename}{Table}
\par
\begin{center}
\begin{tabular}{llll}
&  &  &  \\[-0.3cm] 
Reference (year) & Ising & $q=3$ & $q=4$ \\ 
&  &  &  \\[-0.35cm] \hline
&  &  &  \\[-0.3cm] 
This work & 2.1565(7) & 2.198(2) & 2.290(3) \\ 
\cite{Nightingale} ~(2000) & 2.1667(5) &  &  \\ 
~~\cite{Zhang99}$^1$ (1999) & 2.153(2) & 2.191(6) &  \\ 
\cite{Review}$^2$ (1998) & 2.153(4) & 2.203(11) &  \\ 
\cite{Review}$^3$ (1998) & 2.16(2) & 2.14(3) &  \\ 
\cite{Zheng98} ~(1998) & 2.137(8) &  &  \\ 
\cite{Wang97} ~(1997) & 2.166(7) &  &  \\ 
~~\cite{Sch97}$^4$ (1997) & 2.155(3) & 2.196(8) &  \\ 
\cite{Grassberger} ~(1995) & 2.172(6) &  &  \\ 
\cite{Wang95} ~(1995) & 2.16(4) &  &  \\ 
\cite{Ito93} ~(1993) & 2.165(10) &  &  \\ 
\cite{Jan92} ~(1992) & 2.16(2) &  &  \\ 
\cite{Bonfim} ~(1987) & 2.16(5) & 2.16(4) & 2.18(3) \\ 
\cite{Arcangelis} ~(1986) &  & 2.43(15) & 2.36(20)
\end{tabular}
\end{center}
\par
$^1$ 2D TP lattice. \newline
$^2$ from scaling collapse (Table 3). \newline
$^3$ applying Eq.(\ref{binder1}). \newline
$^4$ from HB algorithm, while $z=2.137(11) \,{\rm and}\, z=2.198(13)$,
respectively for Ising and $q=3$ model from Metropolis algorithm. 
\end{table}


\newpage
\cleardoublepage

\begin{figure}[b]
\begin{center}
\begin{minipage}[t]{0.95\textwidth}
\centering
\includegraphics[width=0.72\textwidth]{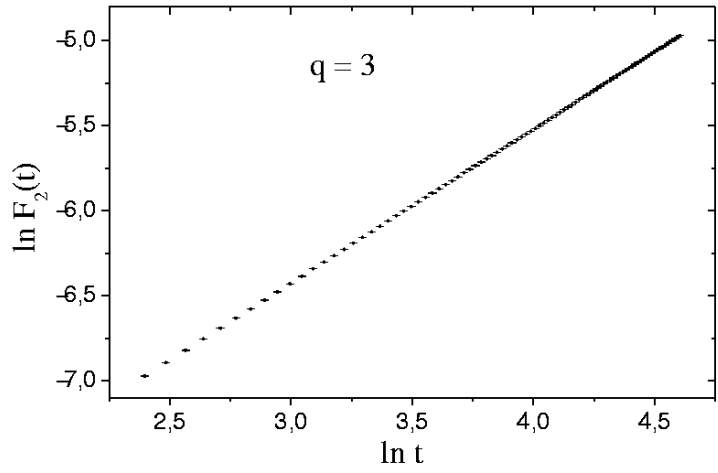}
\renewcommand{\figurename}{(Fig.1)}
\caption{Time evolution of $F_2(t)$ for 2D three-state Potts model.}
\label{Fig. 1}
\end{minipage}
\end{center}
\end{figure}


\begin{figure}[b]
\begin{center}
\begin{minipage}[t]{0.95\textwidth}
\centering
\includegraphics[width=0.72\textwidth]{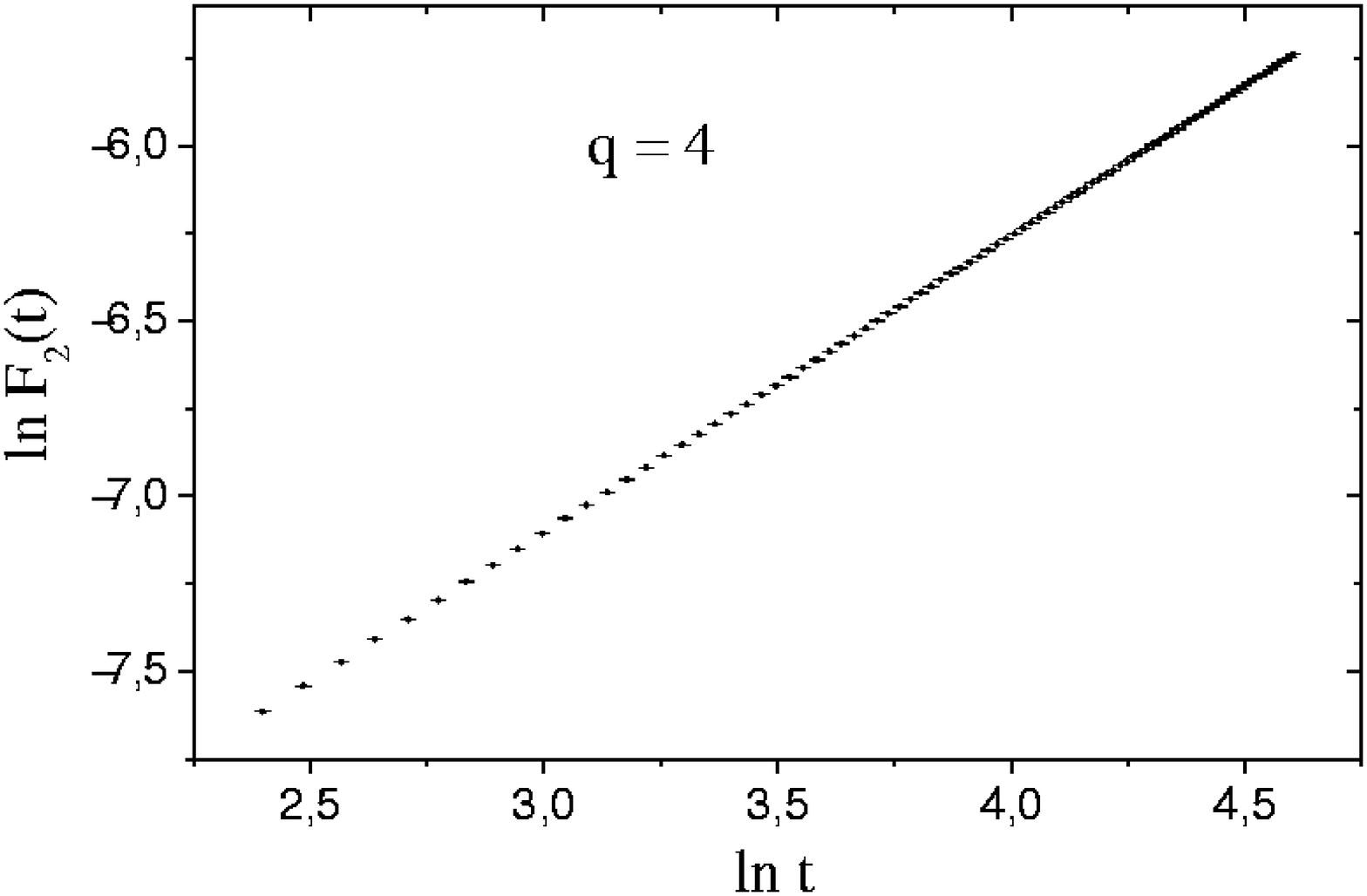}
\renewcommand{\figurename}{(Fig.2)}
\caption{Time evolution of $F_2(t)$ for 2D four-state Potts model.}
\label{Fig. 2}
\end{minipage}
\end{center}
\end{figure}

\end{document}